# Validation of the copper equation of state via shock loading experiments of loosely associated powders

**Author:** Yufeng Wang, Long Hao, Lixin Liu, Fengchao Wu, Shijia Ye, Yuanchao Gan, Yi Sun, Huayun Geng*

**Affiliations:**
National Key Laboratory of Shock Wave and Detonation Physics, Institute of Fluid Physics, CAEP, Mianyang 621900, People's Republic of China

**Corresponding authors:** s102genghy@caep.cn

**Abstract:**
High-fidelity shock experiments were performed on copper powders with controlled porosity via improved target fabrication and assembly. Optical velocimetry and multi-channel pyrometry were used to obtain Hugoniot data, isentropic release paths, and interface temperature histories. The results validate a modified two-phase equation of state (EOS) for copper based on the framework of Greeff et al. The measured Hugoniot shows good agreement with the present model but exhibits significant softening above ~156 GPa relative to the original Greeff EOS, indicating that reduction in lattice specific heat becomes essential when shock temperatures exceed three times the melting point ($T > 3T_m$). Unloading behavior matches hydrodynamic simulations incorporating the recalibrated EOS, confirming its accuracy for off-Hugoniot states. Theoretical analysis of temperature release profiles suggests that the thermal conductivity of shocked copper powders may be considerably higher than first-principles predictions. Crucially, despite heterogeneity in shock heating, the macroscopic dynamic response of copper powders with a porosity of ~1.7 is well captured by an average-density EOS model, supporting the use of porous material experiments for EOS validation under extreme conditions.

**I. Introduction**

Copper is a fundamental material for both dynamic and static high-pressure experiments. Beyond its experimental applications, copper is also widely employed in civilian and military industries, including electronics, aerospace, and defense systems. Consequently, establishing an accurate equation of state (EOS) for copper is critical for advancing experimental research and hydrocode simulations. Over the past five decades, significant efforts have been dedicated to investigating the EOS model and dynamic responses of copper under shock loading,[1-10] substantially improving our understanding of its behavior under extreme pressure and temperature conditions. Notably, advancements in large-scale laser facilities and pulsed power machines have enabled researchers to achieve unprecedented levels of pressure—up to terapascals—through adiabatic or quasi-isentropic compression.[11-12] Such capabilities have opened new avenues for exploring the dynamic properties of copper at conditions that were previously inaccessible. Building upon these experimental findings and complementing them with first-principles calculations, numerous semi-empirical EOS models for copper have been developed.[13-19] These models are indispensable for analyzing experimental data and validating numerical simulations. However, a critical limitation persists: most experimental data are concentrated along specific thermodynamic paths, such as the principal Hugoniot and isotherms. This concentration result in insufficient experimental constraints for off-Hugoniot thermodynamic states and paths. Consequently, while semi-empirical EOS models offer valuable insights, their accuracy in predicting copper's behavior within less-explored regions remains to be validated. Addressing this challenge, future research requires expanding the experimental database to encompass a broader range of thermodynamic conditions. Future research should focus on designing experiments targeting underexplored regions of the phase diagram and leveraging emerging technologies to probe copper's properties at extreme states with higher precision. Furthermore, integrating machine learning techniques with traditional modeling approaches presents a promising avenue for enhancing the predictive capability of EOS models across wider thermodynamic domains.[20] In summary, although significant progress has been made in understanding copper's EOS and dynamic behavior under shock loading, significant gaps remain in characterizing its properties across diverse thermodynamic states, particularly paths beyond the principle Hugoniot.

Shock compression of porous samples provides a unique approach to accessing high-pressure and high-temperature states that lie beyond the principal Hugoniot.[21] Early studies, such as the Los alamos scientific laboratory (LASL) data in the Marsh compendium, predominantly used shorting pins rather than velocity interferometry for Hugoniot measurements. Significant data scatter was observed due to the inconsistent triggering of shorting pins, the use of explosively driven shocks rather than gas guns, and inadequate control of initial sample density. Recent advances in high-precision machining and addictive manufacturing now permit precise control over pore morphology and distribution.[22-24] These capabilities are augmented by laser velocity interferometry, which supersedes shorting pins through nanosecond-resolution tracking of shock wave structures. Combined with rigorous uncertainty quantification methods,[22] these innovations significantly reduce measurement scatter and data uncertainty. Multi-probe diagnostics—including optical pyrometry[25], line-imaging velocity interferometer system for any reflector (VISAR)[26], and time-resolved dynamical X-ray diffraction (DXRD)[27]—collectively deliver pressure, density, temperature, and crystal structure measurements during shock compression, yielding comprehensive characterization of porous materials under dynamic loading. Crucially, phase contrast imaging (PCI)[28-29], DXRD and line-imaging VISAR have made it possible to study the pore compaction mechanism, real-time lattice deformations, and the phase transition kinetics of porous materials at the microscopic/mesoscopic scales. This synergy between sample fabrication/ assembly methods and advanced diagnostics has the potential to transform the paradigm of studying dynamic behaviors of porous materials under shock loading.

Despite advancements in porous material shock compression methodologies, critical knowledge gaps persist. First, pyrometry for shock temperature measurement, though established for solid materials, demands rigorous validation for porous systems regarding its capacity to measure true thermal equilibrium states and reliably validate theoretical models. Multi-channel pyrometry data from shocked porous materials frequently deviate from thermal equilibrium[30-31], limiting their utility in quantitative analysis. While Goodwin et al.[25] first demonstrate the feasibility of time-resolved pyrometry for porous metal oxide under shock compression, their work lacked experiment details and theoretical analysis. Second, there are doubts whether EOS models neglecting micro/mesoscale morphological effects can adequately describe dynamic responses of porous material under shock compression. For instance, Dolgoborodov et al.[32] observed an anomalous reloading/unloading behavior in porous nickel in the middle pressure range (20-35 GPa) that deviated from both mirror-line approximations and hydrodynamic simulations based on macro-homogeneous EOS, and they attribute this phenomenon to heterogeneous melting of shocked porous materials. These unsolved questions underscore the continued need for further investigations into shock dynamics of porous materials.

Based on the above analyses, we performed shock compression experiments on loosely compacted copper powders using optical velocimetry and multi-channel pyrometry. High-fidelity data—including Hugoniot curves, isentropic unloading paths, and interface temperature profiles—were acquired to validate a modified two-phase EOS model (adapted from Greeff et al.[14]) and to examine deviations from EOS predictions, specifically probing non-equilibrium or heterogeneous effects in shocked copper powders.

## II. Experiment

### A. Hugoniot experiment

Planar shock compression experiments were conducted at the Institute of Fluid Physics, China, using a two-stage light gas gun with a 30-mm-diameter bore and a maximum projectile velocity of 7.0 km/s. The copper powders used in these experiments was obtained from University of Science and Technology Beijing. Material specifications indicated a nominal purity of 99.99% and a particle size distribution of 200–1000 mesh (nominal equivalent: 15–75 μm).

The experimental configuration is illustrated in Fig. 1(a). Tantalum (Ta) was selected for both flyers and basements, each with a nominal thickness of 1.5 mm. The flyers had a diameter of 25 mm, while the basements measured 40 mm in diameter. The copper powders and window were encapsulated within a cylindrical cavity (Φ12 mm × 8 mm) positioned between the basement and the stainless-steel vessel. Lithium fluoride (LiF) or sapphire windows were employed to maintain optical transparency under intense shock loading. High-precision machining ensured a surface flatness of less than 5 μm across all components, achieving tight dimensional tolerances. An ultra-thin aluminum foil (~8 μm thick) was bonded to the front surface of the window in direct contact with the copper powders. This foil served to preserve the integrity of the mirrored surface during shock events.

The assembly procedure was conducted as follows. First, a LiF window (Φ12 mm×5 mm) was epoxy-bonded to the bottom surface of the cylindrical cavity within the steel vessel. The residual volume of the cavity (~Φ12 mm×3 mm) was then carefully measured to determine the required mass of copper powders. Copper powders were subsequently packed into the cavity. The loose packing volume slightly exceeded the residual cavity volume to eliminate void space. Next, a tantalum (Ta) basement was installed and mechanically pressed against copper powders, ensuring direct contact between the basement, copper powders, LiF window, and steel vessel. Following assembly, displacements of the basement's front surface relative to initial positions (measured prior to powder filling) were recorded. These displacement measurements enabled quantitative evaluation of uncertainties associated with the filling density of the copper powders.

In Hugoniot experiments, the flyer velocity was measured with <0.5% uncertainty using an optical beam blocking system. Simultaneously, a 21-channel Doppler probe system (DPS) recorded the arrival times of shockwaves and apparent particle velocity profiles at the basement rear surface and the powder/window interface, as shown in Fig. 1(b). Eight DPS probes were positioned at the Φ19 mm and Φ15 mm circumferences, respectively, to monitor shock-wave arrival times at the rear basement interface. Five additional DPS probes at the Φ5 mm circumference and center point recorded shock-wave arrival times and particle velocity profiles at the powder/window interface. Shock transit time through the copper powders was determined via quadratic interpolation of the shockwave front data,[33-34] while the interface velocities were calculated using refractive-index corrections with established accuracy.[35]

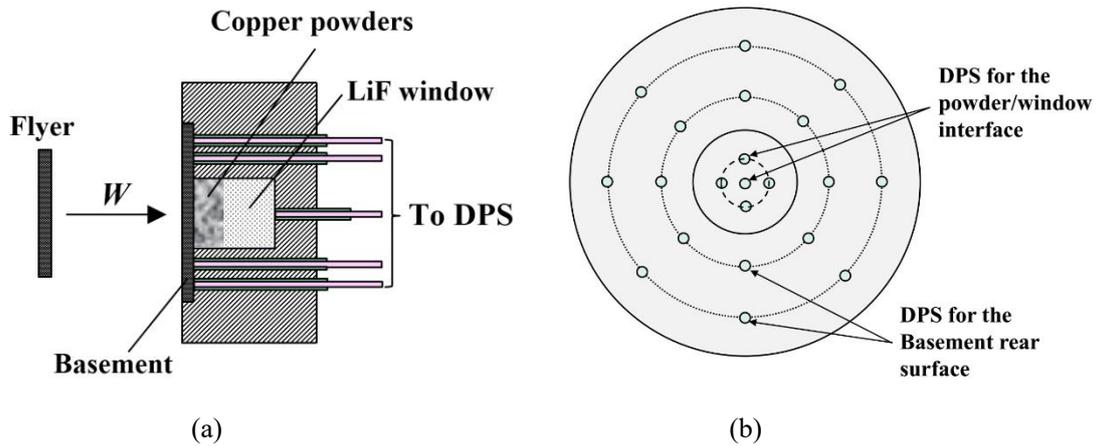

(a)  (b)

Figure 1. Schematic of Hugoniot experiments setup (a) and the arrangement of DPS diagnostics (b). Five DPS probes for powder/window interface are located at the Φ5 mm circumference and center point, and sixteen DPS probes for the basement rear surface are located at Φ19 mm and Φ15 mm circumferences.

**B. Shock temperature experiment**

Figure 2(a) shows the schematic for thermal emission measurements. This configuration closely resembles Figure 1(a) but incorporates a larger single-crystal LiF window (Φ30 mm×8 mm), and the sample chamber (Φ12 mm × 3 mm) was directly machined into the window. Critically, no metallic Al film was attached to the side of the LiF window in contact with the copper powder sample, thereby preventing complications in the interpretation of interface temperature measurements. Interface temperature between the copper powders and LiF window were measured using an optical pyrometer with 16 channels in the wavelength range of 400–800 nm. Each channel incorporated a narrow band width filter of ~10 nm and achieved ~1 ns temporal resolution. Thermal radiation emitted from the interface was collected via optical fibers and transmitted to the pyrometer. The system utilized photomultiplier tubes (PMTs) that converted optical signals to voltage outputs, recorded by oscilloscopes proportional to emission intensity.

Figure 2(b) shows the experimental setup for shock temperature measurements. The DPS probes monitoring shock arrival at the rear substrate interface maintain identical positions to those in Figure 1(b). An optical pyrometer at the center measures interface temperature, while a single DPS probe positioned ~5 mm from the center records shock arrival at the sample/substrate interface.

Before carrying out the experiments, each channel of the pyrometer was calibrated by the spectral radiance from a tungsten-halogen (WBr) standard lamp. Outputs of the pyrometer were recorded by digital oscilloscopes with a sampling rate of ~1.5 GSa/s. Assuming that detect efficiency of PMTs in the pyrometer was the same for transient and quasi-static pulse illumination and that its intensity response was linear with pulse intensity, the measured voltages were converted into light flux based on their calibration, and the interface temperature was then obtained by fitting the measured spectral intensities to Planck distribution for gray-body radiation.[36]

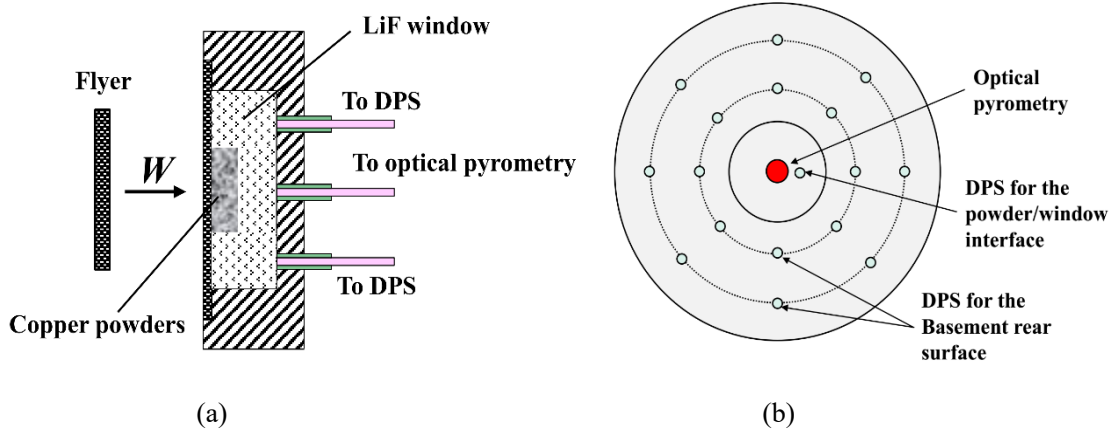

(a)　　　　　　　　　　　　　　　(b)

Figure 2. Schematic diagram for the shock temperature experiment (a) and the arrangement of DPS and optical pyrometry. DPS probes for the basement rear surface are the same with Figure 1(b), Multi-channel optical pyrometer at the center measures interface temperature, while a single DPS positioned ~5 mm from the center records shock arrival at the sample/substrate interface.

### III. EOS modeling of copper

In order to achieve a better description of the liquid phase of copper, we modified the original standard EOS of copper proposed by Greeff *et al.*[14] The model and parameter calibration method are illustrated as follows.

#### a. Solid phase

The Helmholtz free energy for copper in solid phase can be written as

$$F^s(\rho, T) = \phi_0^s(\rho) + F_{\text{ion}}^s(\rho, T) + F_{\text{elec}}^s(\rho, T), \tag{1}$$

where $\phi_0^s(\rho)$ is the static lattice energy at zero temperature, which is also called the cold energy; $F_{\text{ion}}^s(\rho, T)$ is the contribution of lattice vibrations, and $F_{\text{elec}}^s(\rho, T)$ is the contribution of electric excitations.

The cold energy contribution to free energy can be described by the Vinet equation,

$$\begin{cases} \phi_0^s(\rho) = \frac{4B_{0K}}{\rho_{0K}(B'_{0K}-1)^2}[1-(1+X)e^{-X}], \\ X = \frac{3}{2}(B'_{0K}-1)\left[\left(\frac{\rho_{0K}}{\rho}\right)^{1/3}-1\right] \end{cases} \quad (2)$$

where $\rho_{0K}$, $B_{0K}$, and $B'_{0K}$ are the specific volume, the bulk modulus, and the pressure derivative of bulk modulus at zero kelvin, respectively.

The free energy of lattice vibrations in quasi-harmonic approximation is characterized by the Debye model, which is

$$\begin{cases} F_{\text{ion}}^s(\rho,T) = R_m\left[\frac{9}{8}\theta_D + 3T\ln(1-e^{-\theta_D/T}) - TD\left(\frac{\theta_D}{T}\right)\right], \\ D(x) = \frac{3}{x^3}\int_0^x \frac{z^3 dz}{e^z-1} \end{cases} \quad (3)$$

where $R_m$ is the specific gas constant, and $\theta_D$ is the Debye temperature. Assuming that the Grüneisen parameter is only a quadratic polynomial function of the density,[37-38] which correspond to

$$\gamma(\rho) = \begin{cases} \gamma_\infty + g_1\left(\frac{\rho_0}{\rho}\right) + g_2\left(\frac{\rho_0}{\rho}\right)^2, & \rho \geq \rho_0 \\ \gamma_0 + g_3\left(\frac{\rho}{\rho_0}\right) + g_4\left(\frac{\rho}{\rho_0}\right)^2, & \rho < \rho_0 \end{cases} \quad (4)$$

where $\rho_0$ is the density at standard pressure and temperature, then the Debye temperature can be derived as

$$\frac{\theta_D(\rho)}{\theta_{D0}} = \begin{cases} \left(\frac{\rho}{\rho_0}\right)^{\gamma_\infty} \exp\left\{g_1\left[1-\left(\frac{\rho_0}{\rho}\right)\right] + \frac{g_2}{2}\left[1-\left(\frac{\rho_0}{\rho}\right)^2\right]\right\}, & \rho \geq \rho_0. \\ \left(\frac{\rho}{\rho_0}\right)^{\gamma_0} \exp\left\{g_3\left[\left(\frac{\rho}{\rho_0}\right)-1\right] + \frac{g_4}{2}\left[\left(\frac{\rho}{\rho_0}\right)^2-1\right]\right\}, & \rho < \rho_0. \end{cases} \quad (5)$$

where $\gamma_\infty$, $\gamma_0$, $g_1$, $g_2$, $g_3$, and $g_4$ are fitting parameters of data obtained by high pressure experiments, and $\theta_{D0}$ is the Debye temperature at ambient conditions.

The free energy of electric excitations is represented by

$$F_{\text{elec}}^s(\rho,T) = -\frac{\beta_0}{2}\left(\frac{\rho_0}{\rho}\right)^{\Gamma_e} T^2, \quad (6)$$

where $\beta_0$ is the coefficient of electronic heat capacity, and $\Gamma_e$ is an analog of the Grüneisen parameter for electrons. The values of $\beta_0$ and $\Gamma_e$ can be determined by fitting the calculated electronic density of states using *ab initio* methods.

b.  **Liquid phase**

Based on the monatomic liquid model originally developed by Chisolm *et al.*[39], the Helmholtz free energy of copper in liquid phase is

$$F^l(\rho,T) = \phi_0^l(\rho) + F_{\text{ion}}^l(\rho,T) + \Delta F_{\text{ion}}^{l,1} + \Delta F_{\text{ion}}^{l,2} + \Delta F_{\text{elec}}^l \quad (7)$$

where $F_{\text{ion}}^l$ is identical to the previously calculated lattice vibrational free energy contribution for the solid phase. $\Delta F_{\text{ion}}^{l,1}$ is the first correction term of ion free energy due to the entropy difference between solid and liquid at fixed volume, and the value is generally $0.8R_m$ for normal metal like copper based on the liquid vibration-transit theory. Therefore, $\Delta F_{\text{ion}}^{l,1}$ is written as

$$\Delta F_{\text{ion}}^{l,1}(\rho, T) = -0.8 R_m T, \tag{8}$$

$\Delta F_{\text{ion}}^{l,2}$ is the second correction term to ion free energy due to the variation of the specific heat capacity at constant specific volume with temperature. Assuming that the specific heat is only a monotonic decreasing function of temperature, expressed as

$$C_V(T) = C_V(T_m) + \frac{3}{2} R_m (\psi^{-\alpha} - 1), \tag{9}$$

where $T_m$ is the melting temperature, $\alpha$ is a positive constant with a suggested value of 0.35, and $\psi = T/T_m$ is a dimensionless variable. The second correction term can be obtained by integration of the heat capacity in Eq. (9) twice with the temperature

$$\Delta F_{\text{ion}}^{l,2}(\rho, T) = \frac{3}{2} R_m T_m \left[ \frac{\psi^{1-\alpha}}{\alpha(1-\alpha)} + \psi \ln \psi - \left(1 + \frac{1}{\alpha}\right)\psi - \frac{\alpha}{1-\alpha} \right], \tag{10}$$

$\Delta F_{\text{elec}}^{l}$ is the contribution of electric excitations in liquid phase, which can be approximated to take the same form as the corresponding solid term, Eq. (6), due to the similarity of electronic density of states in both phases. $\phi_0^l(\rho)$ is the cold energy term of liquid which can be determined by equating the Gibbs free energies in solid and liquid phase along the melting curve following the Lindemann melt rule, and theoretical details have been clearly illustrated by Chisolm et al.[40]

Consistent with the approach of Greeff et al., we determined the EOS parameters using the experimental constrains including the 300 K isotherm, principle Hugoniot, and melting curve. The calculated EOS parameters of copper in solid phase are presented in Table I. For the liquid phase, the monatomic liquid model has indicated that the contribution of the lattice vibrational and electric excitation contributions to the EOS are approximately equivalent to their solid-phase counterparts, meaning that $F_{\text{ion}}^{l}(\rho, T) = F_{\text{ion}}^{s}(\rho, T)$ and $F_{\text{elec}}^{l}(\rho, T) = F_{\text{elec}}^{s}(\rho, T)$.[40] Consequently, differences in Helmholtz free energy between solid and liquid phases manifest primarily in the cold energy and two correction terms of ion free energy including $\Delta F_{\text{ion}}^{l,1}$ and $\Delta F_{\text{ion}}^{l,2}$ arising from atomic transit motion between random valleys. Furthermore, the copper EOS model in this work maintains consistency with the Greeff model in both framework and key parameter values, with the principal distinction in the liquid phase being our explicit inclusion of $\Delta F_{\text{ion}}^{l,2}$.

**Table I**. EOS parameters of the copper in solid phase.

| $B_{0K}$ (GPa) | $B_{0K}'$ | $\rho_{0K}$ (g/cm³) | $\theta_{D0}$ (K) | $g_1$ | $g_2$ | $g_3$ | $g_4$ | $\beta_0$ (kJ/(g·K²)) | $\Gamma_e$ |
|---|---|---|---|---|---|---|---|---|---|
| 143.70 | 5.25 | 8.9303 | 321.70 | 1.2591 | 0.0747 | 3.4092 | -2.4088 | 1.09E-08 | 0.717 |

## IV. Experiment results and discussion

### a. Hugoniot curves in the *D-u* plane

A summary of the experimental data is shown in Table II. $W$, $\rho_{00}$, and $D$ are measured values of the velocity of Ta flyers, the average density of copper powders, and the shock velocity. $P$ and $u$ are values of the pressure and particle velocity, which can be obtained by the impedance match method. $u_w$ is the experimental value of the sample/LiF interface velocity. For shots 3~7 using sapphire windows, the values of $u_w$ are omitted due to unavailable reflective index corrections under shock loading.

**Table II**. Summary of the results of Hugoniot experiments. The superscripts "a" or "b" of the shot number in the first column stands for the experiment using the LiF or sapphire, respectively.

| Shot | $\rho_{00}$/(g/cm$^3$) | $W$/(km/s) | $u$/(km/s) | $D$/(km/s) | $P$/(GPa) | $u_w$/(km/s) |
|---|---|---|---|---|---|---|
| 1[a] | 5.61±0.06 | 2.600±0.005 | 1.913±0.026 | 4.462±0.076 | 47.89±1.97 | 2.01±0.02 |
| 2[a] | 5.54±0.07 | 5.026±0.012 | 3.486±0.042 | 7.039±0.083 | 135.94±4.96 | 3.86±0.04 |
| 3[b] | 5.56±0.07 | 6.428±0.015 | 4.377±0.053 | 8.383±0.118 | 204.01±7.91 | − |
| 4[b] | 5.55±0.06 | 6.772±0.019 | 4.581±0.058 | 8.832±0.122 | 224.55±8.37 | − |
| 5[b] | 5.57±0.07 | 6.149±0.012 | 4.178±0.068 | 8.277±0.357 | 192.62±13.86 | − |
| 6[b] | 5.59±0.08 | 6.429±0.015 | 4.355±0.077 | 8.514±0.395 | 207.27±16.25 | − |
| 7[b] | 5.57±0.07 | 5.092±0.012 | 3.499±0.045 | 7.307±0.138 | 142.40±6.31 | − |
| 8[a] | 5.45±0.07 | 3.833±0.012 | 2.727±0.043 | 5.864±0.183 | 87.15±5.21 | 2.96±0.03 |

Figures 3 presents Hugoniot data form Table II (Our experiment: magenta and red dots with error bars) alongside the LASL experiment datasets corresponding to different initial porosities of copper (dots without error bars). A number of features are worthy of note. The first is that our results demonstrate reduced scatter and enhanced compressibility compared to the LASL dataset[1] using porous samples with initial densities of ~5.74g/cm$^3$—a consequence of LASL's density fluctuations (5.60-5.98g/cm$^3$ in sintered specimens). Another attractive feature is that our porous Hugoniot curves shows no high-pressure intersection with LASL's ~8.93 and ~7.32 g/cm$^3$ curves, contrasting sharply with LASL's intersecting linear extrapolation for ~5.74 g/cm$^3$ curves. This absence of crossing aligns with Trunin *et al*.[41] who considered such intersections unphysical for porous material like copper, although formal disproof remains elusive.[42-43] The last feature is that our experiments achieve ~225 GPa peak pressure—~90 GPa greater than the maximum pressure of LASL dataset with similar porosity—enabling unprecedented exploration of off-Hugoniot states. To sum up, high-fidelity powder shock methodology yields more reliable, less biased data with substantially extend pressure ranges versus LASL benchmarks.

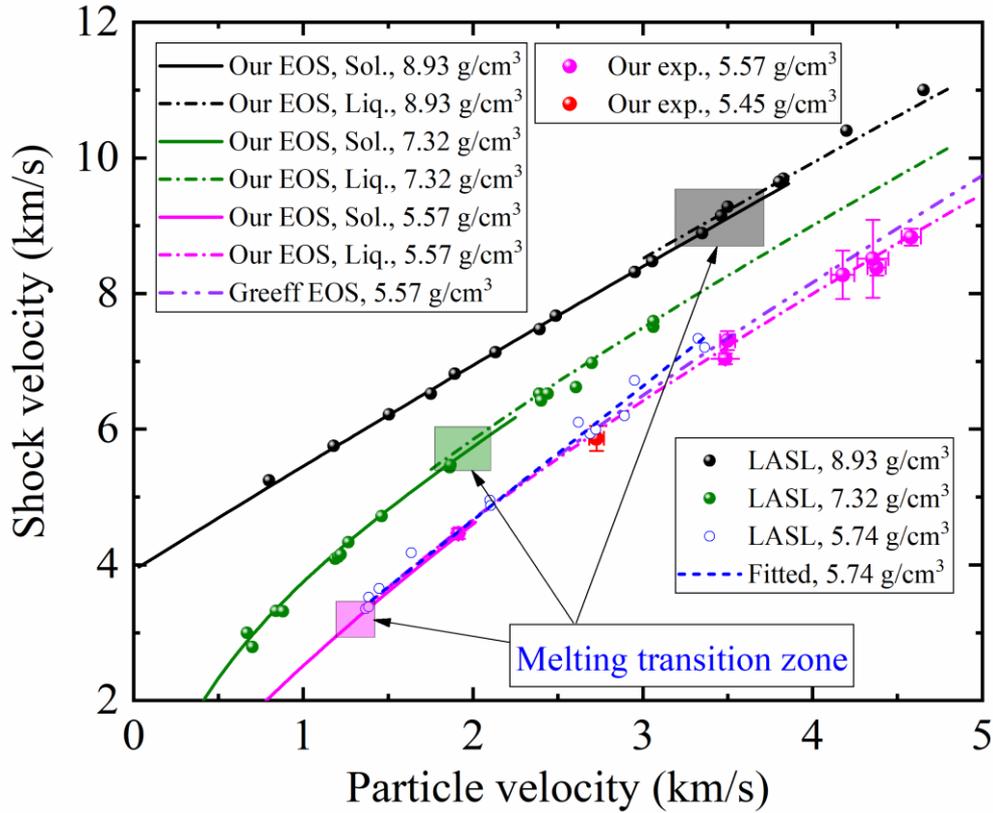

Figure 3. Hugoniot data obtained in our shock experiments (red or pink dots) in comparison with the LASL Hugoniot data (black, green, or hollow blue dots) corresponding to different initial porosities of copper and theoretical Hugoniot curves calculated by our EOS model or the Greeff model. Our experimental data exhibit significant softening compared to the Greeff EOS model and LASL experimental data at particle velocities above 3.0 km/s, with the deviation increasing at higher velocities. After incorporating the modified term proposed in this work, the theoretical results show good agreement with the experimental data.

Figure 3 also plots Hugoniot curves for varying initial copper densities (solid-dashed lines: solid phase; dot-dash lines: liquid phase) calculated by our two-phase EOS model. The model's porous copper Hugoniot predictions align well with most experimental data including both ours and LASL's, excepting LASL's nominal ~5.74 g/cm$^3$ dataset, reinforcing the superior fidelity of our measurements. As an extension of the conclusion by Greeff *et al.*[14], we find that shock melting results in the stiffening of the Hugoniot curves for both solid and porous samples. In order to get a more *detailed* description on the shock melting behavior and its influence on the $D$-$u$ curve, the corresponding theoretical Hugoniot curves of copper with different porosities in the $T$-$P$ phase diagram are illustrated in Figure 4. It is obvious that increasing porosity progressively narrows the melting transition zone, consequently reducing deviations between solid and liquid $D$-$u$ curves within the shock melting regime.

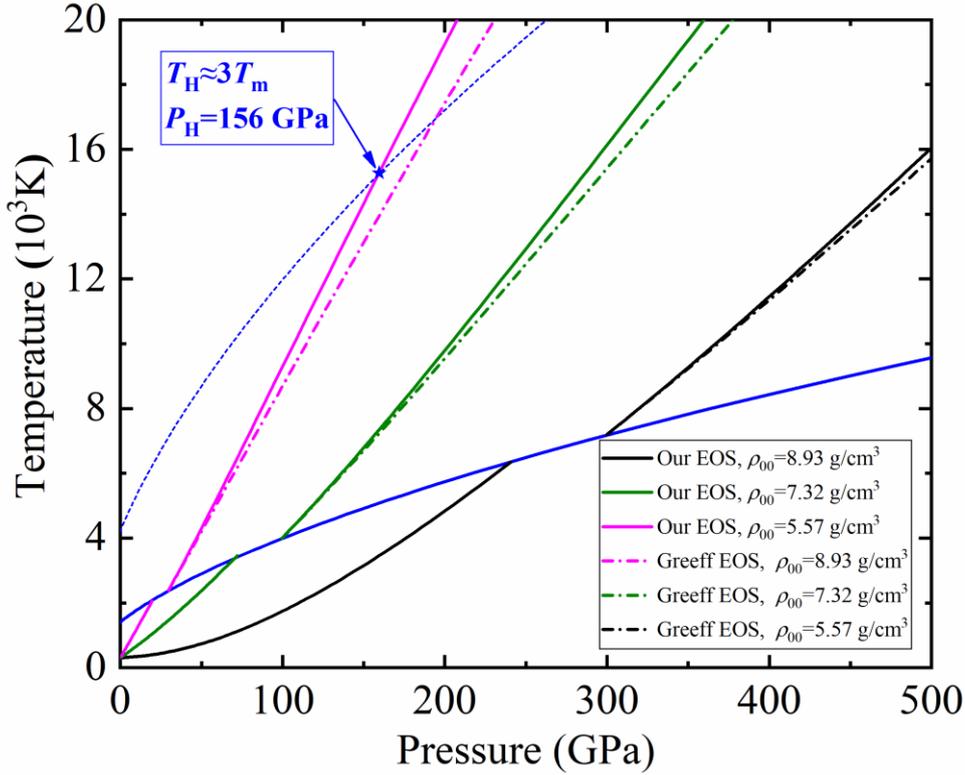

Figure 4. Hugoniot curves for varying initial porosities in the temperature-pressure plane calculated using our two-phase EOS and the Greeff EOS model. The continuous and dotted blue lines represent the melting temperature and threefold melting temperature as a function of pressure, respectively. $T_H$ and $T_m$ denote the shock temperature and melting temperature, respectively, and $P_H$ represents the shock pressure. The intersection of the threefold melting temperature curve and the Hugoniot curve on the $T$–$P$ plane, as calculated using the two-phase EOS model proposed in this work, occurs at $P$ = 156 GPa. At this shock pressure, the predicted shock temperature shows a deviation of nearly 10% from that obtained using the Greeff EOS. This discrepancy grows gradually with increasing shock pressure.

Complementary Hugoniot curves for liquid copper at ~5.74g/cm$^3$ initial density, calculated using the original Greeff EOS model, are plotted as magenta dashed lines in Figure 3 and Figure 4 against our EOS model. Figure 4 demonstrates that the discrepancies between Greeff EOS and ours primarily emerge at elevated temperatures. Since Eq. (9) expresses specific heat as a melting-temperature scaling function, these differences exhibit analogous scaling behavior. Crucially, at $T_H \approx 3T_m$ and $P_H$=156 GPa, shock temperatures show ~10% relative deviation while corresponding shock velocities deviate greater than 2.0% (Figure 3), exceeding acceptable tolerances for precise Hugoniot fitting[38]. Consistently, our two-phase EOS model generates softer high temperature $D$-$u$ curves than Greeff's original model (as also shown in Figure 3), with experimental data validating this behavior. We therefore conclude that high-temperature lattice specific heat reduction inducing Hugoniot softening in both solid and porous copper, establishing metal powders shock experiments as an effective methodology for EOS validation.

b.  **isentropic curves in the *P-u* plane**

Figure 5 displays a representative DPS frequency spectrum from the copper powders/LiF window interface in shot-1. A broad velocity distribution band emerges immediately after the steep rising edge. The band narrows progressively over time, evolving into a relative narrow velocity plateau at ~50 ns that persists stably for approximately 100 ns. Subsequent arrival of the release wave at the interface then initiates particle velocity decay. The post-edge velocity distribution originates from surface roughness at the Cu/LiF contact zone, with the transition time to stable plateau correlating directly to Cu powder characteristic particle size.

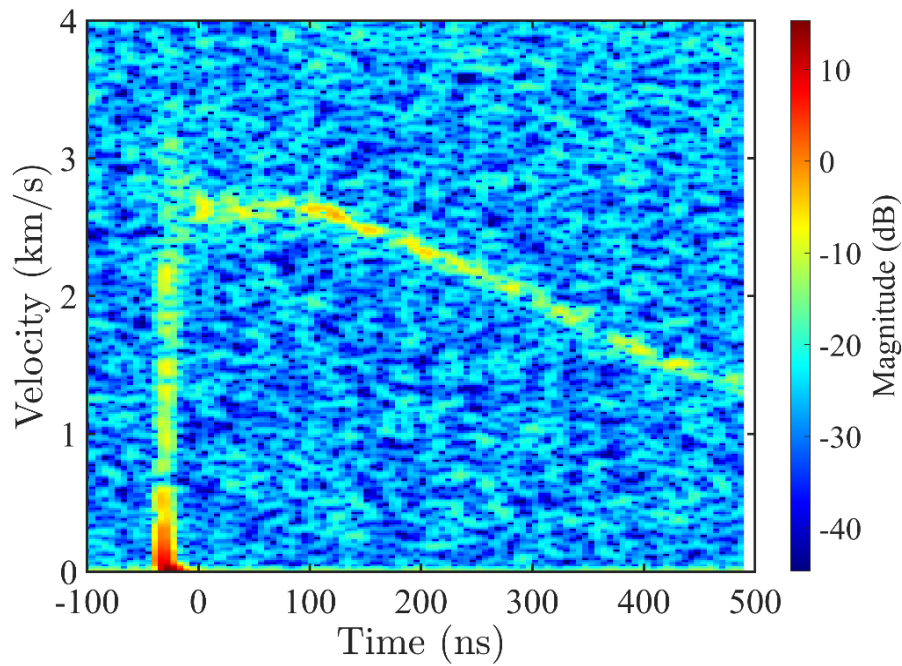

Figure 5. Frequency spectrum of a typical DPS signal at the powder/window interface in shot-1

Figure 6 presents interface particle velocity profiles for shots 1, 2, and 8, measured at the center position by DPS probes and derived from ridge lines of DPS frequency spectra.[44] All profiles show oscillations indicative of the heterogeneous structure of shocked copper powders. Hydrodynamic simulations reproduce both the plateau velocities and the unloading trends observed experimentally. These simulations used a one-dimensional multiphase hydro-dynamic code (1D-MHDC), with the copper powder represented by the two-phase EOS of copper and a Herrmann-type P-α model. Strength effects were neglected due to liquefaction of the powder under high shock pressures. The Ta flyer and LiF window were modeled with a Grüneisen EOS and the Steinberg-Cochran-Guinan constitutive model. The agreement validates the two-phase EOS of copper under conditions extending beyond the Hugoniot curve.

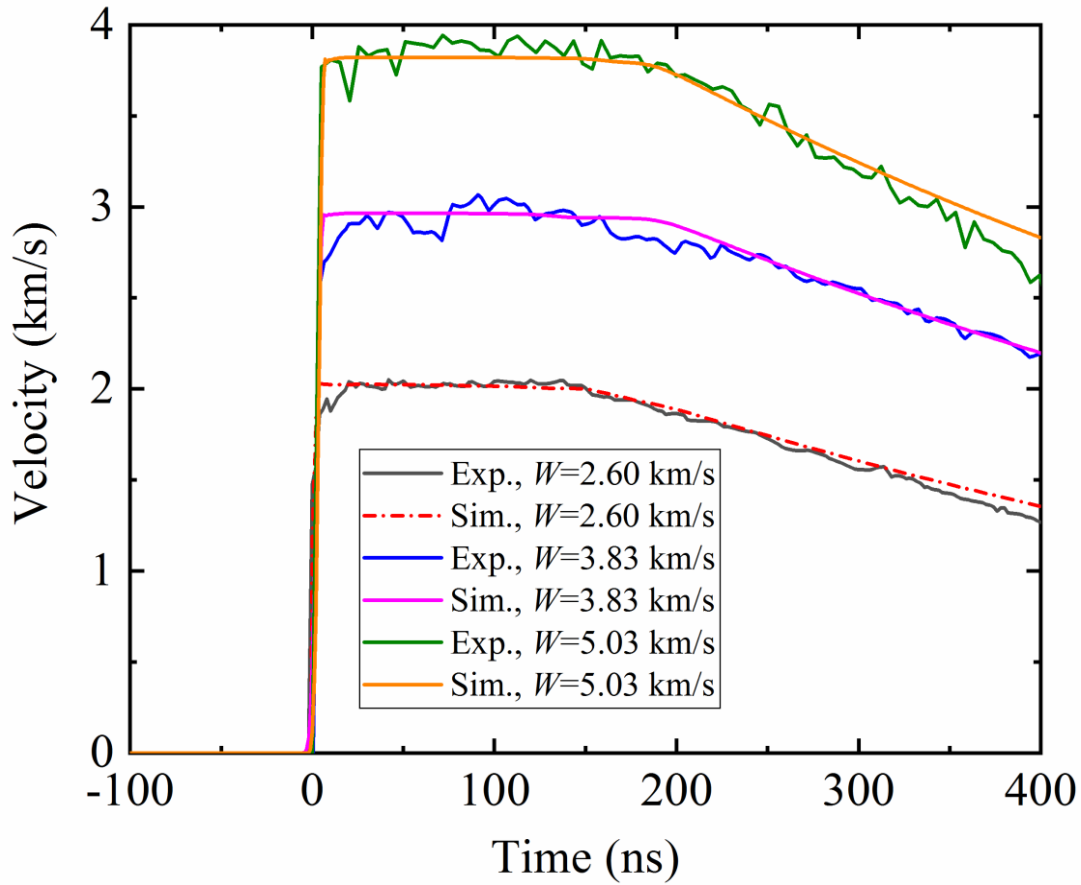

Figure 6. Comparison of interface velocity profiles, measured at the center position by DPS probes in shots 1, 2, and 8, with hydrodynamic simulation results obtained using the 1D-MHDC based on our two-phase EOS model. ANSYS/AUTODYN two-dimensional simulations indicate that the velocity measurement from the central DPS probe begins to be influenced by lateral release waves at approximately 400 ns.

Figure 7 plots Hugoniot states for copper powders (initial density ~5.57g/cm$^3$) for shots 1, 2, and 8 and corresponding release states at the powder/window interface in the $P$-$u$ plane, alongside Hugoniot and mirror curves for both porous and solid copper. The experimental unloading paths connecting Hugoniot and corresponding release states align more closely with mirror-image curve of solid copper's principle Hugoniot rather than with the porous material's Hugoniot curve. This derivation necessitates modification of the classical mirror reflection approximation used in impedance matching for porous materials. Crucially, as shock compressed copper solidify during initial during initial compression and retain solid phase during unloading, it is not surprising that their isentropic release paths naturally conform to solid copper's mirror curves through the first shock states.

While Kobayashi et al.[45] report unloading behaviors consistent with our results, Dolgoborodov et al.[32] observe a different phenomenon: anomalous reloading/unloading isentropes in porous nanosized nickel significantly deviate from both solid and porous Hugoniot mirror curves in the middle pressure range (20-35 GPa), and they attribute this deviation to non-equilibrium melting processes enhanced by material heterogeneity. Although Dolgoborodov's pressure range is much lower than ours, porous nickel melts at ~17 GPa, while porous copper melts at ~40 GPa in our work (Figure 4). Thus, their intermediate pressure range is comparable to our low-pressure experimental data. These observations raise concerns about Dolgoborodov's interpretation. First, the melting phase transition exhibits negligible influence on copper's compressibility in our data (Figure 3). Second, their analysis lacks substantive discussion of window correction uncertainties due to water's complex phase behavior during interface velocity measurements. Resolving these issues requires additional experiments following Dolgoborodov's design to measure unloading behaviors over wider pressure ranges in P-u diagrams, along with molecular dynamics simulations.[46-47]

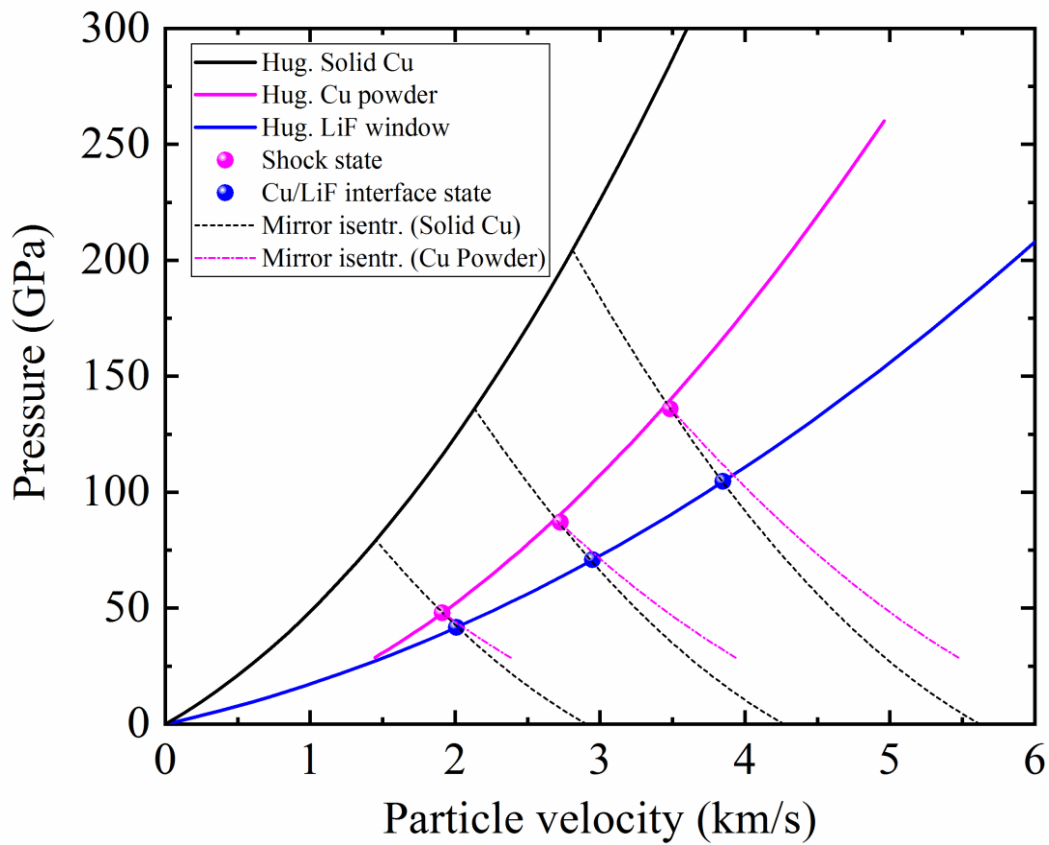

Figure 7. Shock states (red dots) and the corresponding unloading states (blue dots) in the *P-u* plane, and the dashed black (pink) curves are the mirror reflection curves of the solid (porous) Hugoniot curves of copper.

c. **Interface temperature under shock loading and release**

Figure 8 presents time-resolved thermal emission spectra of shot-7 experiment, with measured spectral radiance uncertainty of ~2%. The radiance initiates at point A, marking shock wave arrival at the interface. From A to B, a period corresponding to the duration of the interface particle velocity plateau in Figure 6, a broad peak emerges due to gap-flash and heterogeneous light emission at the powder/window interface, followed by gradual decay reflecting temperature reduction during stress release. C approximates the time at which the lateral release wave begins to affect temperature measurement at the sample's center. Beyond point D—where the shock front reaches the window's free surface—radiance decreases abruptly due to wave interaction induced opacity changes in LiF. Thus data beyond this point are consequently excluded from analysis.[48]

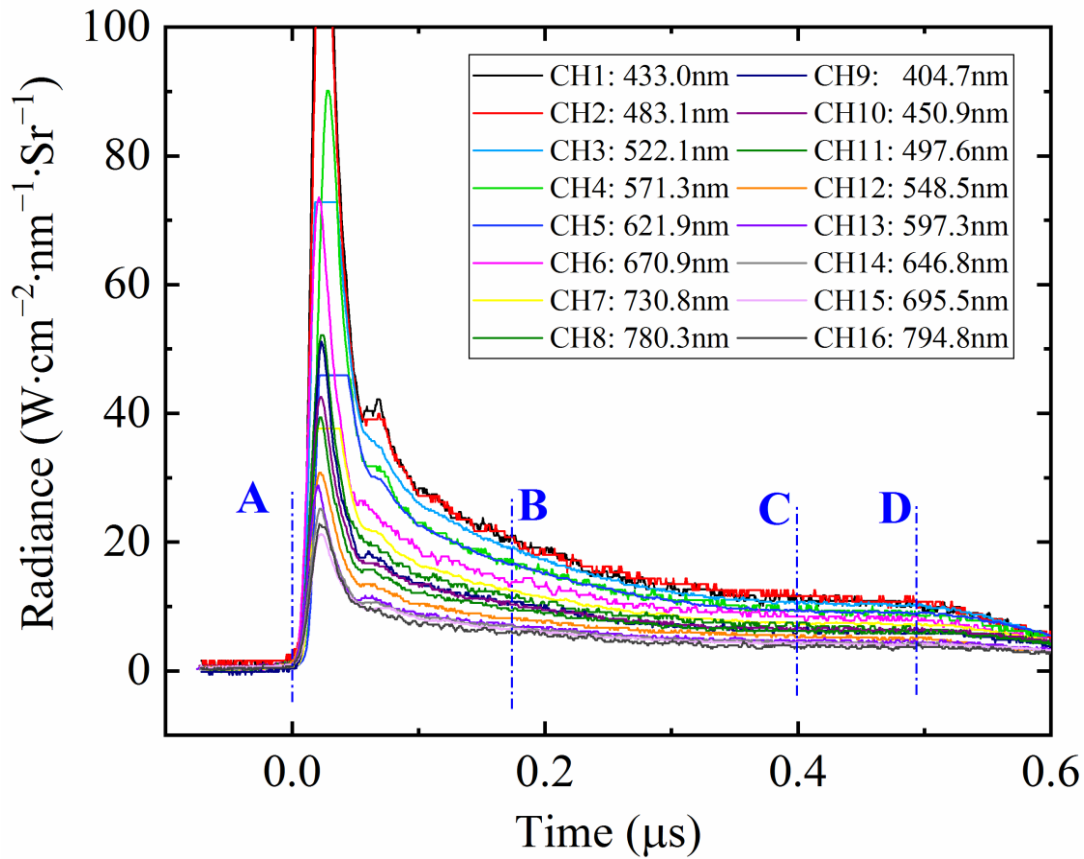

Figure 8. Time-resolved thermal emission spectra measured by the 16-channel pyrometry. A denotes the arrival time of the shock wave at the powder/window interface. B marks the arrival time of the rarefaction wave. C approximates the time at which the lateral release wave begins to affect velocity/temperature measurement at the sample's center. D corresponds to the time when the shock front reaches the free surface of the LiF window.

Figure 9 compares interface temperatures derived from gray-body Planckian fits to spectral intensities with particle velocity profile simulated by 1D-MHDC. The particle velocity profile exhibits a clear single-wave structure, characterized by a stable plateau between points A and B. From B to D, the velocity decreases gradually with an approximately constant slope, although some deviation may occur in the CD segment due to the influence of lateral release waves. In contrast, the temperature profile shows distinct evolutionary features: a sharp peak emerges during the AB interval, followed by a monotonic decrease from B to C owing to the overtaking release wave. The gradually diminishing slope of the temperature curve suggests a transition toward interfacial thermal equilibrium. From C to D, the temperature remains nearly constant, likely resulting from the combined effects of both lateral and overtaking release waves. Based on the above analyses, the thermal response equilibrates with the bulk material around point C. This conclusion is also supported by Mandal et al.'s DXRD experimental results[27] and theoretical estimation based on the thermal diffusion theory[48]. Thus, thermal emission data and temperature profile around point C represent bulk porous copper near thermal equilibrium, enabling direct comparison with theoretical predictions from the two-phase EOS of copper.

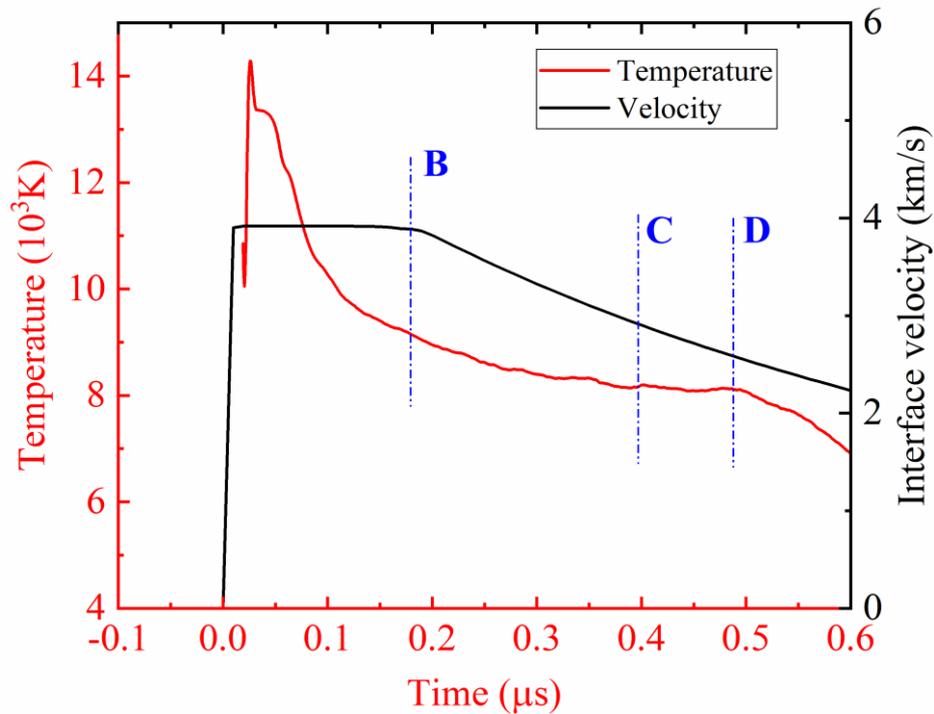

Figure 9. Interface temperature profile and the corresponding simulated velocity profiles in shot-7. Labels B, C, D have the same meaning as those in Figure 8.

Both temperature and velocity measurements in the CD segment may be affected by lateral release waves. This effect cannot be captured by 1D hydrocode. Nevertheless, it remains valuable to compare the experimental interface temperature evolution during the BD segment with results from 1D simulations. Applying impedance matching methods to simulated velocity profiles (Validated against experiments in Section IV.b), we drive interface pressure histories $P(t)$, combining with the experimental interface temperature profile $T(t)$ yields the temperature-stress release path $T(P)$ shown in Figure 10. Theoretically, release temperatures for shocked copper powders and LiF window can derive from our EOS and Zhou, *et al.*[49] respectively. While the Grover-Urtiew model provides the fundamental interface temperature ($T_I$) formulation[50-51]

$$T_I = T_S - \frac{T_S - T_W}{1 + \alpha_{SW}} \quad (11)$$

where $T_S$ and $T_W$ are the bulk temperature of sample and window, respectively, and $\alpha_{SW}$ is the ratio of their respective thermal effusivities

$$\alpha_{SW} = \left(\frac{k_S \rho_S C_{P,S}}{k_W \rho_W C_{P,W}}\right)^{\frac{1}{2}} \quad (12)$$

where $k$ is the thermal conductivity, $\rho$ is the density, and $C_P$ is the constant pressure heat capacity; and the subscripts "S" and "W" stand for sample and the LiF window, respectively. According to equations (11) and (12), accurate interface temperature prediction via the Grover-Urtiew model is limited by uncertain thermal transport parameters. To simplify the calculation of the interface temperature, $\alpha_{SW}$—the thermal effusivity ratio between copper and LiF—was assumed to be a constant. The calculated results show the variation of interface temperature with pressure for different values of $\alpha_{SW}$. It can be observed that larger values of $\alpha_{SW}$ lead to lower temperature profiles. Moreover, the best agreement between experimental data and theoretical prediction is achieved when $\alpha_{SW}$ is around 0.4.

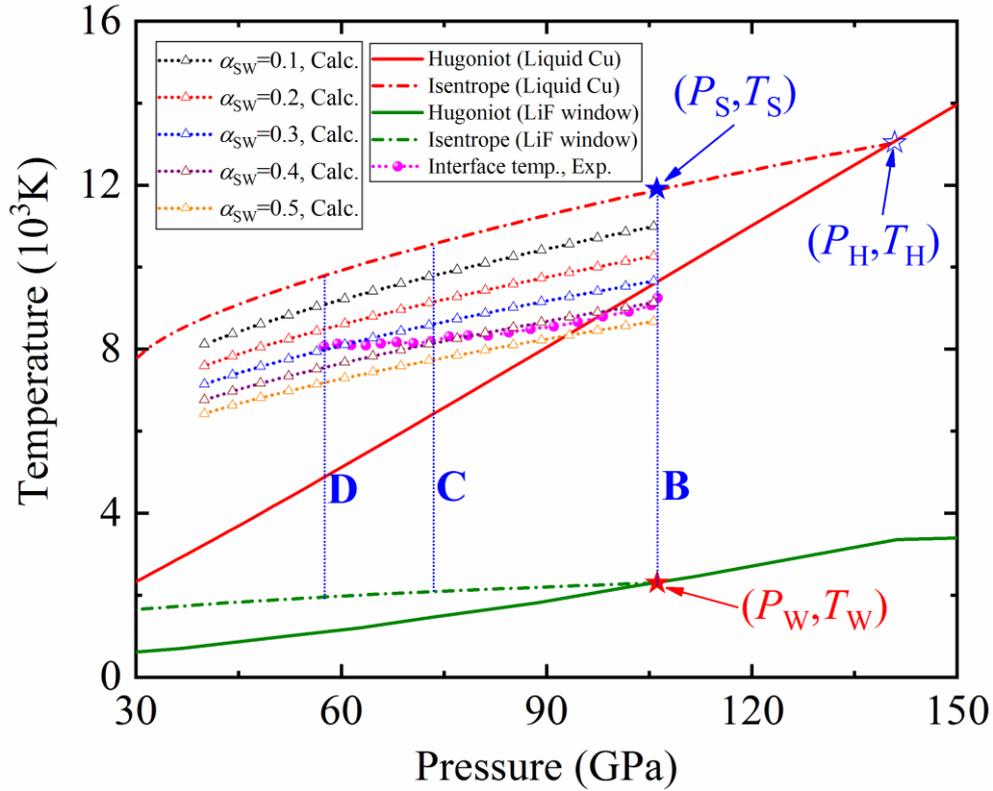

Figure 10. Comparison of the experimental release path in the *T-P* plane in shot-7 and theoretical calculation results based on our two-phase EOS model. Labels B, C, D have the same meaning as those in Figure 8. $P_H$ and $T_H$ denote the shock pressure and shock temperature of the sample, respectively; $P_S$ and $T_S$ represent the pressure and temperature after shock reflection at the powders/window interface; $P_W$ and $T_W$ correspond to the pressure and temperature behind the transmitted shock wave in the window.

First-principles simulations of the thermal conductivity for copper[10] and the LiF window[52] reported in literature indicate a range of 0.1 to 0.25 for $α_{SW}$. Consequently, our theoretically predicted temperature-pressure release paths exceed experimental measurements. However, these *ab initio* calculations neglect heterogeneity effects in shocked copper powders. Gilev[53] observed electrical resistance exceeds first-principles predictions under shock loading—attributed to defect generation from non-equilibrium compression states. By Wiedemann-Franz law, this implies that thermal conductivity of shocked copper powders is significantly reduced compared to theoretical value. This inference provides a plausible pathway to reconcile the discrepancy between the theoretical and experimental temperature-pressure release paths: the reduced thermal conductivity, attributable to microstructural heterogeneity, leads to an α value around 0.4—substantially higher than that predicted for homogeneous materials.

Given the inherent diagnostic challenges for shock temperature experiments—even in homogeneous systems—quantitative utilization of these results for EOS validation or interfacial heat conduction studies remains fundamentally constrained. To improve the reliability of such measurements, two key experimental modifications are proposed: increasing the sample aspect ratio to mitigate the influence of lateral release waves, and increasing the flyer thickness to extend the duration of the shock plateau, thereby allowing more time for the sample to approach thermal equilibrium. Meanwhile, future work must prioritize quantifying heterogeneity effects on electrical conductivity in shock-compressed porous copper while developing advanced theoretical models that integrate heterogeneity and non-thermal equilibrium states, enabled by in situ time-resolved microscopic diagnostics coupled with atomistic simulations.

## V. Summary and conclusions

Shock experiments on copper powders with initial densities of 5.4–5.7 g/cm³ were conducted to validate a modified two-phase EOS based on the framework by Greeff et al. Enhanced target fabrication enabled precise density control, yielding Hugoniot data with lower uncertainty than LASL results at comparable porosity. A peak pressure of ~225 GPa was achieved—approximately 90 GPa above LASL values.

Our Hugoniot data show significant softening compared to both LASL data and the original Greeff EOS. This discrepancy is attributed to reduced lattice specific heat at temperatures above three times the melting point ($T > 3T_m$), an effect not previously considered. A modified two-phase EOS constructed using our data accurately captures the Hugoniot behavior across various initial porosities, consistent with both our and published results.

The model also predicts experimental interface velocity profiles. Agreement in plateau velocities and unloading histories confirms its validity for off-Hugoniot states. Unloading paths in the $P$–$u$ plane align with the mirror-image Hugoniot of solid copper—not porous copper—demonstrating clear loading/unloading asymmetry.

Although the limited shock plateau duration prevented direct measurement of interface equilibrium temperature, theoretical analysis suggests sufficient time may allow thermal equilibrium. The thermal conductivity of shocked copper powders may also exceed first-principles predictions, likely due to microstructural heterogeneity.

Importantly, despite microscale heterogeneity, the macroscopic dynamic response of shocked copper powders with a porosity of ~1.7 is well-described within an EOS framework based on averaged density evolution. This establishes shock experiments on porous materials as a robust complementary approach for EOS validation under extreme conditions.

**ACKNOWLEDGEMENTS**:

The authors are grateful to Q. S. Wang, Q. Kang, Q. W. Fu, T. P. Dou, and H. F. Xian for their assistance with experiments. This work was supported by the National Key R&D Program of China (Grant No. 2021YFB38002300), the Foundation of the State Key Laboratory of Shock Wave and Detonation Physics (Grant No. 2023JCJQLB05401), and the National Natural Science Foundation of China (Grant No. 12272363&12202418).